%% file: article.tex
\documentclass[a4paper,fleqn,usenatbib]{mnras}
\usepackage{newtxtext,newtxmath}
\usepackage[T1]{fontenc}
\usepackage{ae,aecompl}

\usepackage{graphicx}   % Including figure files
\newcommand{\JHK}{JHK_{\rm s}}
\newcommand{\Ks}{K_{\mathrm{s}}}
\newcommand{\MK}{M_{K_{\mathrm{s}}}}

\title[Carbon-rich Miras in the Galactic bulge]
{Discovery of carbon-rich Miras in the Galactic bulge}
\author[N. Matsunaga~{et~al.}]
{Noriyuki Matsunaga$^{1}$\thanks{E-mail:matsunaga@astron.s.u-tokyo.ac.jp},
John W. Menzies$^{2}$,
Michael W. Feast$^{3,2}$,
\and
Patricia A. Whitelock$^{2,3}$,
Hiroki Onozato$^{4}$,
Sudhanshu Barway$^{2}$,
\and
and Elias Aydi$^{2,3}$
\\
$^{1}$ Department of Astronomy, The University of Tokyo, 7-3-1 Hongo, Bunkyo-ku, Tokyo 113-0033, Japan \\
$^{2}$ South African Astronomical Observatory, PO Box 9, Observatory 7935, South Africa \\
$^{3}$ Astronomy Department, University of Cape Town, Rondebosch 7701, South Africa\\
$^{4}$ Astronomical Institute, Graduate School of Science, Tohoku University, 6-3 Aramaki Aoba, Aoba-ku, Sendai, Miyagi 980-0857, Japan
}

\date{Accepted 2017 May 15. Received 2017 May 15; in original form 2017 Feb 28}

\pubyear{2017}

\begin{document}
\label{firstpage}
\pagerange{\pageref{firstpage}--\pageref{lastpage}} \pubyear{2017}
\maketitle

\begin{abstract}
Only one carbon-rich (C-rich, hereinafter)
Mira variable has so far been suggested as a member
of the Galactic bulge and this is in a symbiotic system.
Here we describe a method for selecting C-rich candidates from
an infrared colour-colour diagram,
$(J-\Ks)$ vs $([9]-[18])$. Follow-up low-resolution spectroscopy resulted
in the detection of 8 C-rich Mira variables from a sample of
36 candidates towards the Galactic bulge.
Our near-infrared photometry indicates that two of these,
including the known symbiotic, are closer than the main body of the bulge
while a third is a known foreground object.
Of the 5 bulge members, one shows \ion{He}{i} and [\ion{O}{ii}]
% He\,{\sevensize I} and [O\,{\sevensize II}]
emission and is possibly another symbiotic star. Our method
is useful for identifying
rare C-rich stars in the Galactic bulge and elsewhere.
The age of these C-rich stars and the evolutionary process which
produced them remain uncertain. They could be old
and the products of either binary mass transfer or mergers,
i.e.\  the descendants of blue stragglers, 
but we cannot rule out the possibility that they belong to
a small {\it in-situ} population of
metal-poor intermediate age ($< 5$~Gyr) stars in the bulge
or that they have been accreted from a dwarf galaxy.
\end{abstract}


\section{Introduction}
\label{sec:Intro}

Work in recent years on the Galactic bulge has shown that it is complex,
consisting of stars with a range of chemical compositions, ages and
distributions \citep[e.g.,][]{Ness-2016,McWilliam-2016,Catchpole-2016}.
It is striking that although Mira variables are common in the bulge
only one carbon-rich Mira has been suggested as 
a possible bulge member, and that is in a symbiotic system
(Miszalski, Miko{\l}ajewska, \& Udalski \citeyear{Miszalski-2013}).
However, a small number of giant carbon-rich stars have been found in
the Galactic bulge \citep{Azzopardi-1991, Tyson-1991}
and their nature will be considered below.

Miras are long period ($ P \geq 100$ days), large amplitude variables
lying at the top of the asymptotic giant branch (AGB).
Kinematic and other studies \citep{Feast-2000,Feast-2008,Menzies-2011}
indicate that the Mira pulsation period is a good indicator of age
and initial mass, and theoretical evolutionary models of single low to
intermediate mass stars are consistent with this \citep{Vassiliadis-1993}.
Miras are separated into two major groups according to their surface chemistry:
oxygen-rich and carbon-rich (hereinafter, O-rich and C-rich, respectively).
In addition to CO molecules which are present in the atmospheres of stars
in both groups, O-rich stars, which have C/O$<1$, show molecules such as
H$_2$O, TiO and SiO whereas C-rich stars, with C/O$>1$, show molecules such
as C$_2$ and CN.
Stars born from material of normal composition have more oxygen than carbon,
but thermal pulses in shell-burning AGB stars can dredge up carbon
produced in the interior to the surface, thereby making them C-rich.
The ages and initial metallicities of stars will determine whether
they become C-rich or not, and this third dredge-up path will produce
C-rich stars aged at around 0.5--5~Gyr \citep{Mouhcine-2003,Marigo-2008}.

As regards C-rich stars generally, several other methods of production
are possible. The giant and subgiant C-rich CH stars and Barium stars are
in long period binary systems and are believed to have atmospheres polluted
by a companion which has passed through the AGB phase
\citep[][and references therein]{McClure-1990}. Cool dwarf C-rich stars
have been found in significant numbers. Although they are not all
now in binary systems, they must have been paired with a much more
massive star at some stage in their evolution
\citep{Green-1994,Totten-2000,Green-2013,Plant-2016}.
Various paths for the formation of CEMP stars (carbon-enriched metal poor
stars) are discussed by \cite{Beers-2005}.
Binary interaction has been suggested for R type stars \citep{Izzard-2007}
and for the Azzopardi bulge C-rich stars
(Azzopardi, Lequeux, \& Rebeirot \citeyear{Azzopardi-1988};
\citealt{Whitelock-1993}; \citealt{Ng-1997}),
while a binary merge has been suggested  for a C-rich Mira in
a globular cluster (Feast, Menzies, \& Whitelock \citeyear{Feast-2013}). 

\citet{Cole-2002} assumed that stars in the bulge
with $(J-\Ks)_0 > 2$ were C-rich.
However, because mass-losing O-rich AGB stars can also be very red
\citep[e.g.,][]{Wood-1998,Ojha-2007},
particularly in metal-rich environments, this is clearly not
a sufficient condition for identifying C-rich stars
in the bulge \citep{Uttenthaler-2015,Catchpole-2016}.
The present paper reports the discovery of C-rich Miras in the bulge based
on a new method using infrared colours to select candidates and
also on follow-up spectroscopy.
We also briefly discuss the nature of these stars and their possible
relationship to the Azzopardi C-rich stars.

\section{Spectroscopic Observations}
\label{sec:Observation}

\subsection{Target selection}
\label{sec:selection}

We use the $(J-\Ks)$ and $([9]-[18])$ colours to identify 
candidate C-rich stars \citep{Ishihara-2010}.
C-rich and O-rich stars, especially those with thick circumstellar dust shells,
tend to show different colour trends in these bands.
As summarised by \citet{Ita-2010}, dusty C-rich stars often show
the SiC emission band at {11.3~\micron}  within the [9] band
in addition to a continuum excess due to amorphous carbon dust,
while dusty O-rich stars have the silicate bands at 
9.8 and {18~\micron}  \citep{Lancon-2000,Loidl-2001,Wright-2009}.
In the near-infrared bands, C-rich AGB stars have
strong C$_2$ and CN bands in $J$, whilst there are no correspondingly
strong bands in $\Ks$ which, unlike the $K$ band, is minimally affected by
the strong CO band longward of {2.29~\micron}.
In contrast, for O-rich AGB stars, strong H$_2$O absorption has a
large impact on both $J$ and $\Ks$ magnitudes while
TiO and VO bands affect $J$.
These features at least give a qualitative explanation for
the different locations of C-rich and O-rich Miras.

\begin{figure}
\begin{minipage}{83mm}
\begin{center}
\includegraphics[clip,width=0.95\hsize]{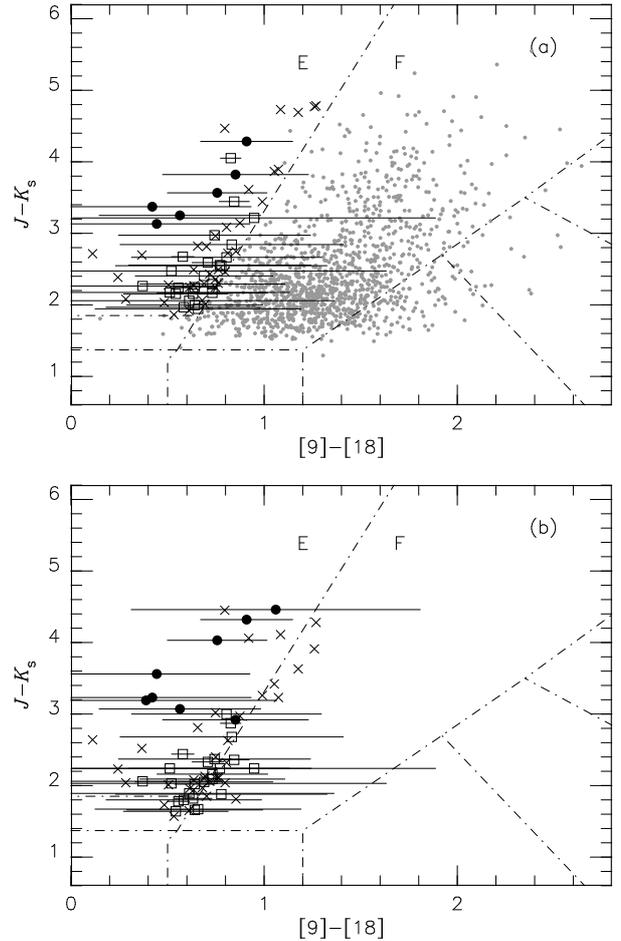}
\end{center}
\caption{
Colour-colour diagram, $(J-\Ks)$ vs $([9]-[18])$, for OGLE-III Miras in the bulge
indicated by grey dots and for our spectroscopic targets.
Those found to be C-rich and
O-rich are indicated by filled circles and open squares, respectively,
while crosses indicate those that were not observed or were unclassified.
Error bars for $[9]-[18]$ colours are given for targets
that we describe in this work.
Panels (a) and (b) use 2MASS and IRSF near-infrared colours,
respectively, except for three objects whose $J-K$ colours are taken
from \citet{Catchpole-2016} in panel (b).
The regions E and F are defined by \citet{Ishihara-2011}
and host mainly C-rich and O-rich stars, respectively.
\label{fig:CCD}}
\end{minipage}
\end{figure}

As the primary list of Miras towards the bulge we used
the catalogue by \citet{Soszynski-2013}
which contains 6528 Miras, together with
more than 230,000 other types of long-period variable
as a result of the third phase of Optical Gravitational Lensing Experiment
(OGLE).
The near- and mid-infrared colours are taken from the 2MASS
\citep{Skrutskie-2006} and \textit{AKARI}
\citep{Ishihara-2010} surveys, respectively.
Among the OGLE-III Miras, we found that 4148 objects have both of
the required colours, and we identified 66 candidates that fall
in the region expected for C-rich stars (Fig.~\ref{fig:CCD}, region E).
We observed 33 of them, listed in Table~\ref{tab:spec}, as we describe below.

We added a few targets from the catalogue of \citet{Catchpole-2016},
which is based on $JHKL$ photometry (our numbers~06, 07, and 33). 
We combined their $J-K$ colours (without system transformation) with 
the \textit{AKARI} $[9]-[18]$ colours from the catalog by \citet{Ishihara-2010}
for identifying them as C-rich candidates. 
Object No.~06 (IRAS~17446$-$4048) was already
identified as C-rich
based on its IRAS/LRS spectrum 
(Groenewegen, de Jong, \& Baas \citeyear{Groenewegen-1993}); it is also
listed in the General Catalog of Galactic Carbon stars \citep{Alksnis-2001}.

When we cross correlate the variable catalogues
\citep{Soszynski-2013,Catchpole-2016} with the photometric ones
(2MASS and \textit{AKARI}), we used a relatively large radius, $5\arcsec$,
in order to avoid missing any C-rich stars whose positions got
somehow disturbed by error.
Nevertheless, every target we observed has good positional matches in
the 2MASS and \textit{AKARI} catalogues within $2.5\arcsec$ (mostly
within $0.5\arcsec$ of a 2MASS counterpart and within $1.5\arcsec$ of
an \textit{AKARI} counterpart).
Miras almost always dominate other nearby stars in the infrared
and the cross-correlation with infrared catalogues is straightforward.

It should be noted that the interstellar reddening can affect
our target selection.
Although the OGLE-III survey regions (selected for their $I$-band survey)
are expected to be less affected by the interstellar reddening
than the Galactic mid-plane, in some regions $A_V$ reaches
6~mag \citep[see e.g.][]{Nataf-2016b},
which corresponds to ${\sim}0.8$~mag in $E_{J-\Ks}$.
Considering that the effect of the reddening is larger
in the near infrared than in the mid infrared,
this may give offsets on the colour-colour diagram which bring some objects
near the boundary from the region F (for O-rich) to E (for C-rich).
We have not corrected the photometry of the selected candidates for
interstellar reddening (in Fig.~\ref{fig:CCD})
although this might possibly be done in an approximate way
using nearby giants. Again, the priority of our target selection was
to include as many C-rich candidates as possible.

\subsection{Observations and data analysis}

For this programme we used the SpUpNIC spectrograph \citep{Crause-2016}
on the SAAO 1.9-m telescope at Sutherland. Because the Miras are all very red,
we installed the GR8 grating with a resolving power of 1200 to cover
the wavelength range from 5600 to 9200~\r{A}. With a slit width corresponding
to 2.1~arcsec, this gave spectra with 
a resolution of about 8~\r{A}. Conditions were rarely photometric throughout
the one week's observing run, 13--19 July, 2016,
and the Moon was near full, resulting often in high sky backgrounds.

The spectra were bias-corrected and flat-fielded in the usual way, and
IRAF routines were used to calibrate the spectra in wavelength.
For some objects, close companions coupled with relatively poor seeing
made extraction of the target spectra difficult.  
While the SpUpNIC usually reads out images
with 2 pixels binned (in the spatial direction),
we observed the object No.~05 without this binning
to maximize chance of separating it from a much brighter M-type companion.
No attempt has been made to flux-correct the spectra.

\input{tab1.tex}

\subsection{Results}

Spectra for 32 targets (Table~\ref{tab:spec}) were of sufficient
quality for further analysis.
That for No.~8 is of low quality but a tentative classification seems
plausible.
A further three targets (Nos.~34, 35, and 36) were either too faint
or too badly blended to be useful,
so they are not included in the following discussion.

\begin{figure}
\begin{minipage}{83mm}
\begin{center}
\includegraphics[clip,width=0.95\hsize]{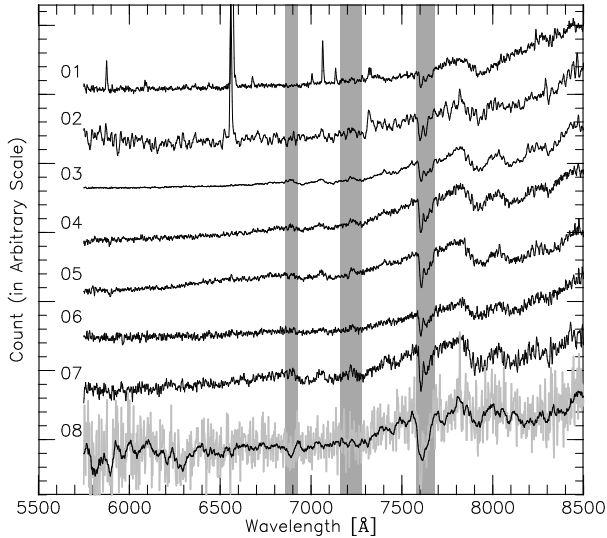}
\end{center}
\caption{
Spectra of C-rich Miras.
The telluric absorption features, the strongest of which are due to
O$_2$ 6867~\r{A} (B band), H$_2$O 7164~\r{A} and O$_2$ 7594~\r{A} (A band),
are shown as shaded vertical bands.
The spectrum of No.~8 has a rather low signal-to-noise ratio
as illustrated by the grey curve, but the black curve after smoothing
shows the similarity to the other spectra.
\label{fig:specC}}
\end{minipage}
\end{figure}

\begin{figure}
\begin{minipage}{83mm}
\begin{center}
\includegraphics[clip,width=0.95\hsize]{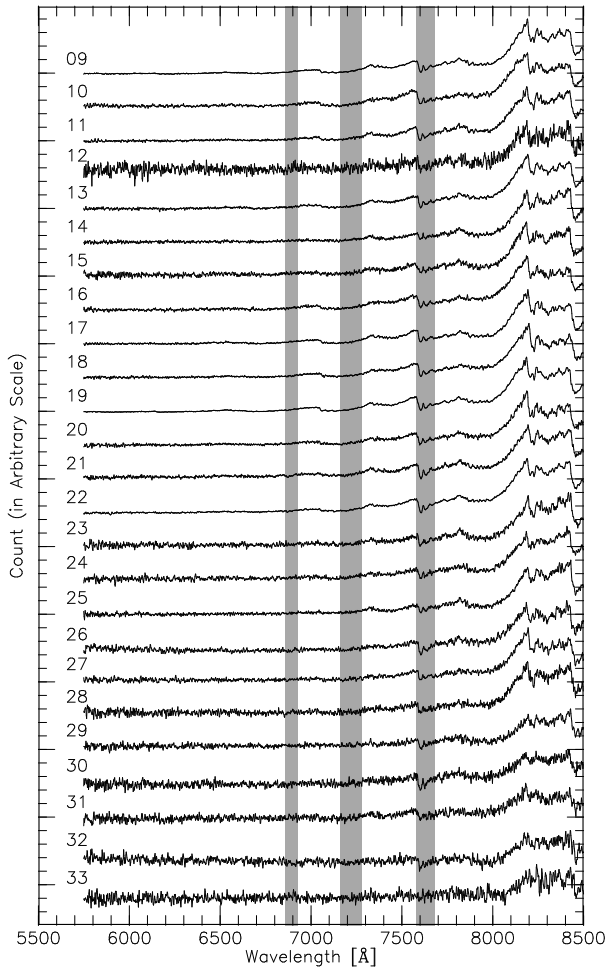}
\end{center}
\caption{
Same as Fig.~\ref{fig:specC} but for O-rich Miras.
\label{fig:specM}}
\end{minipage}
\end{figure}

Figs.~\ref{fig:specC} and \ref{fig:specM} show
the spectra for C-rich and O-rich Miras respectively.
The spectra in Fig.~\ref{fig:specC} clearly show
the sequential vibrational bands of CN, 2--0, 3--1, and 4--2,
between 7850 and 8400~\r{A} \citep{Wyckoff-1970}. 
No.~08 has a relatively bright close companion that,
together with seeing and bright sky background, made it difficult to
extract the spectrum, resulting in a poor signal-to-noise ratio.
There is no evidence of TiO bands and it is probably a C-rich star
based on the appearance of the spectrum in the region around the CN bands. 
The spectra in Fig.~\ref{fig:specM}, in contrast,
are characterized by strong TiO bands at 8194 and 8430~\r{A}
together with TiO bands longward of 7667~\r{A}
beside the telluric A band at around 7594~\r{A}
\citep{Wyckoff-1970,Lancon-2000}.
We thus gave the classification as listed in Table~\ref{tab:spec},
seven C-rich, one probably C-rich and 25 O-rich Miras.

In Fig.~\ref{fig:specC}, our spectrum of No.~01 is similar to that
shown by \citet{Miszalski-2013} who identify the star (H1-45) as a symbiotic.
The spectrum of No.~02, though rather weak, shows emission lines from 
H${\rm \alpha}$, He\,{\sevensize I} (7065~\r{A})
and [O\,{\sevensize II}] (7319$+$7330~\r{A}) as in that of No.~1.
The He and [O\,{\sevensize II}] lines in particular 
suggest a binary system, but there are
no suitable diagnostic high excitation lines in this part
of the spectrum to allow us to classify it definitely as symbiotic.
The Raman bands at 6825 and 7082~\r{A} are absent from both spectra,
but this probably indicates that the companion star does not have
a sufficiently high temperature to excite them (see the discussion
in \citealt{Miszalski-2013}).

\section{Follow-up photometry and distances}
\label{sec:IRSF}

\subsection{New photometric data}
\label{sec:phot}

\begin{figure*}
\begin{minipage}{160mm}
\begin{center}
\includegraphics[clip,width=0.80\hsize]{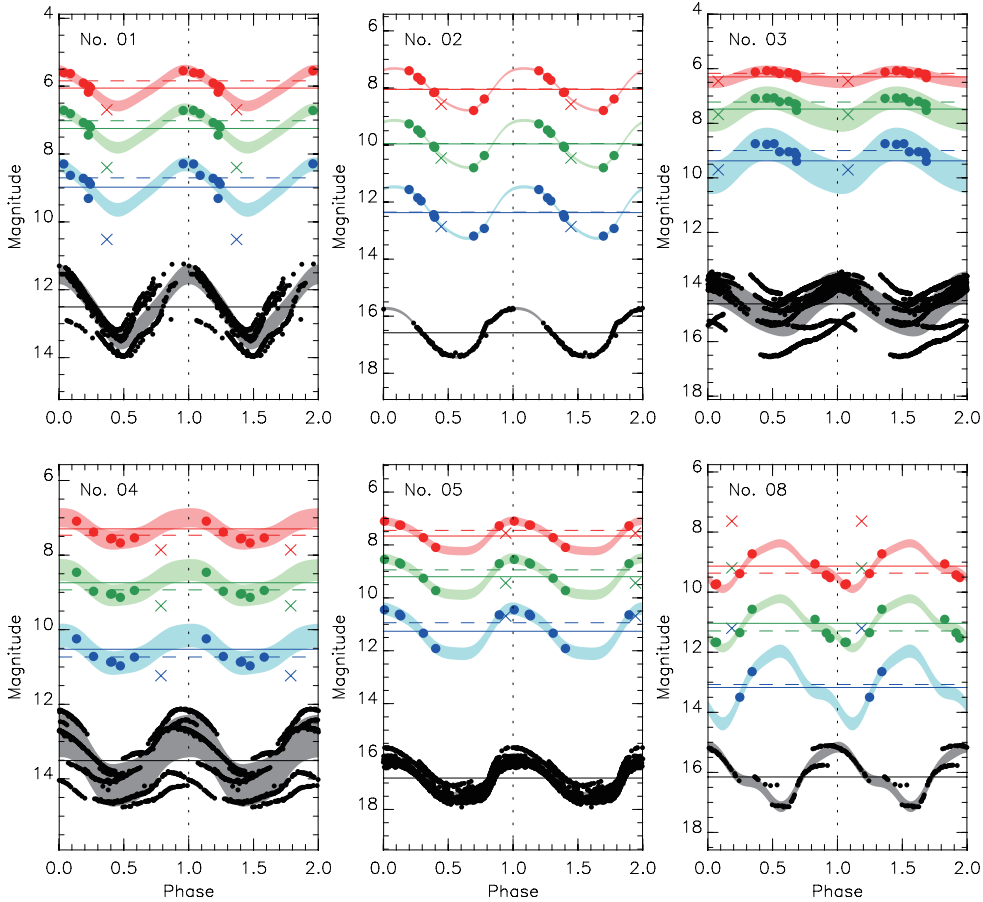}
\end{center}
\caption{
Phased light curves of C-rich Miras.
Black, blue, green, and red points indicate photometric measurements
in $I, J, H$, and $\Ks$. The cross symbols mark 2MASS magnitudes.
The fitted light curves are drawn for each band. The width of a $I$-band curve
corresponds to the standard deviation of the photometric points 
around the fitted curve, while the widths of the curves for other bands
correspond to that of the $I$ band but scaled according to
the difference in amplitudes. The filled and dashed horizontal lines indicate
the mean magnitude for a fitted curve and the simple average of
the photometric points, respectively.
\label{fig:LCs2_C}}
\end{minipage}
\end{figure*}

We used the 1.4-m Infrared Survey Facility at Sutherland,
with the SIRIUS near-infrared camera \citep{Nagashima-1999,Nagayama-2003},
to estimate mean magnitudes of our targets.
We observed each target for 6--9 epochs between 2015 June and 2016 August.
For No.~36, we obtained 18 measurements in each band,
but we used two different integration times and so there are 
pairs of measurements close in time at each epoch.
Fig.~\ref{fig:LCs2_C} plots phased light curves of C-rich Miras in $\JHK$
in addition to the ones in the OGLE $I$ band. 
The third-order Fourier series,
\begin{equation}
m(t)=m_0 + \sum_{i=1}^{3} A_i \cos ~(2\upi it/P+\phi_i) 
\end{equation}
was fitted for the $I$ band, and it was adjusted to fit the light curves
in $\JHK$ with changing the mean ($m_0$), total amplitude, and maximum phase. 
The amplitudes of our C-rich objects are larger than 1~mag, which
is typical for C-rich Miras \citep{Whitelock-2006}.
The derived means are compared with 
the simple averages of $N$ measurements in Table~\ref{tab:phot}.
The 2MASS magnitudes, crosses in Fig.~\ref{fig:LCs2_C}, were not considered
when the amplitudes were scaled and tend to be offset from the fitted curves
(see the next paragraph).
Although our photometric data show relatively large gaps in phase,
the light curves seem to give reasonable estimates of mean.
For the six C-rich Miras in Fig.~\ref{fig:LCs2_C},
the difference between the two kinds of mean is estimated to be
approximately $-0.1 (\pm 0.2)$~mag in each of the $\JHK$. 
This suggests that our estimates of mean are as good as 0.2~mag.
The mean magnitudes from the fitted light curves are used in the following, 
but the difference between two kinds of means does not affect the conclusions.
For the three objects from \citet{Catchpole-2016},
we adopt the published $JHK$ 
magnitudes which are averages of two measurements in most cases.

\input{tab2.tex}

\begin{figure}
\begin{minipage}{83mm}
\begin{center}
\includegraphics[clip,width=0.95\hsize]{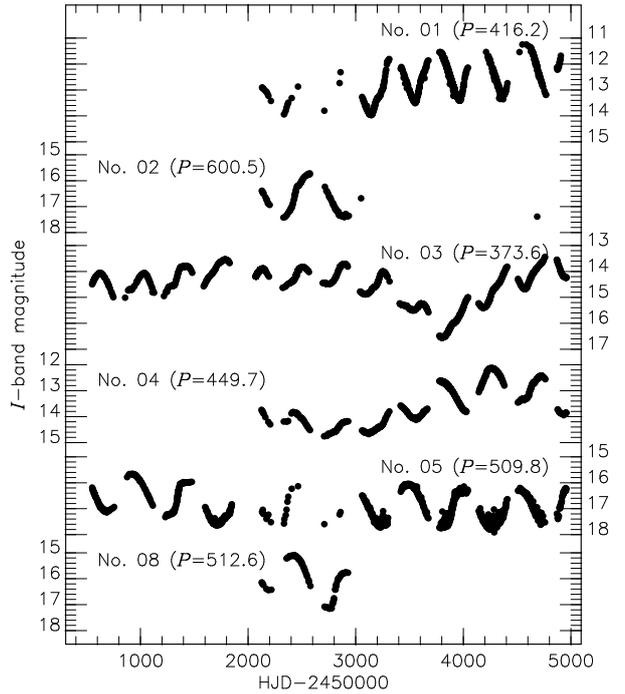}
\end{center}
\caption{
OGLE-III $I$-band light curves of C-rich Miras.
\label{fig:LCs_C}}
\end{minipage}
\end{figure}

It is worthwhile to give some further comments on the light curves
of our C-rich Miras.
The OGLE-III $I$-band light curves of the C-rich Miras
are plotted in Fig.~\ref{fig:LCs_C} when available.
All of them show cycle-to-cycle variations except 
No.~02 for which the time coverage is very short.
It is known that C-rich Miras tend to show such variations
more often than O-rich ones \citep{Whitelock-2003,Whitelock-2006}.
The fact that the 2MASS points in Fig.~\ref{fig:LCs2_C} are not
aligned with the IRSF light curves may be explained by such variations
since the 2MASS photometry was obtained over 15~years ago.

For two objects, Nos.~03 and 08, the fitted curves show large offsets
in phase between the $I$-band and $\JHK$-band ones in Fig.~\ref{fig:LCs2_C}.
The long time interval between the OGLE-III datasets and
our photometry, ${\sim}10$~years, makes it difficult to discuss such offsets,
they can be understood as follows.
No.~03 shows a large fluctuation in the $I$-band light curve and
some cycles seem to show unphased trends. 
The OGLE-III light curve of No.~08 covers less than two cycles
(Fig.~\ref{fig:LCs_C}) and seems to show a variation between the two minima.
The period and phase may have significant errors which produce
the inconsistency in phase between the two datasets.

As presented in Fig.~\ref{fig:CCD}, all eight C-rich stars
fall in the region E although the large errors in $[9]-[18]$ make
it hard to be sure of their precise locations in this diagram.
On the other hand, many of the candidates are found to be actually O-rich,
especially among those near the boundary between the regions E and F
(Fig.~\ref{fig:CCD}). In fact, our follow-up photometry suggests that
there were biases in the 2MASS $J-\Ks$ colours
that we used for the targets selection.
This may be due to colour variations during a cycle; the 2MASS magnitudes
were single-epoch values.
Some O-rich stars could have been included in our C-rich candidates
due to the interstellar reddening as mentioned in Section~\ref{sec:selection}.

It should be pointed out that there are clearly O-rich stars
with $J-\Ks$ colours significantly larger than 2~mag.
There are such red stars among our spectroscopic O-rich samples,
and a number of objects among the OGLE-III bulge Miras
have similarly large $J-\Ks$ and are located in the region F.
These are probably O-rich objects with
thick circumstellar dust shells.
The $J-\Ks$ colour is not sufficient to select C-rich stars.

\subsection{Colour-magnitude diagram}

\begin{figure}
\begin{minipage}{83mm}
\begin{center}
\includegraphics[clip,width=0.95\hsize]{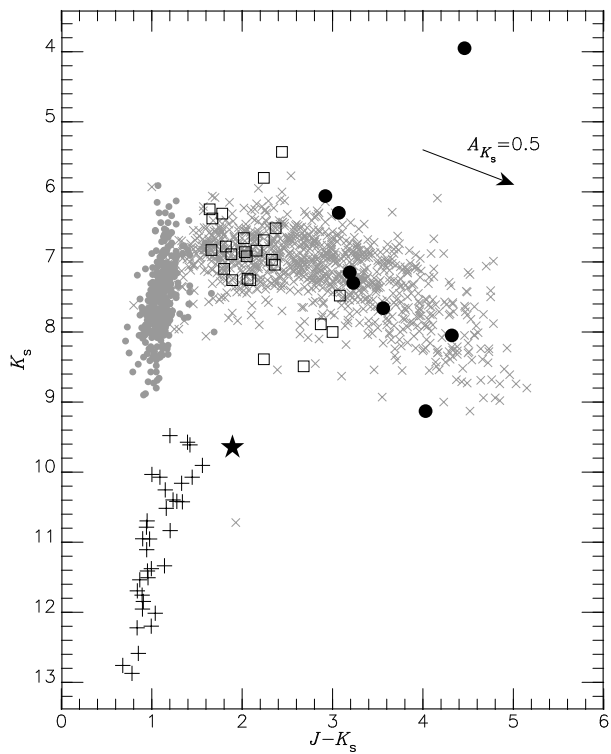}
\end{center}
\caption{
Colour-magnitude diagram.
Filled circles indicate C-rich Miras among our targets
while we found that the Miras indicated by open squares are O-rich.
The plus and star symbols illustrate 2MASS photometry for C-rich stars 
reported by \citet{Azzopardi-1991} and by \citet{Hynes-2014}, respectively.
Grey dots and crosses indicate OGLE-III Miras in the LMC
which are classified as O-rich and C-rich, respectively,
by \citet{Soszynski-2009}; their 2MASS $\Ks$ magnitudes are offset
by 4~mag correcting for the distance difference between
the bulge and the LMC.
\label{fig:JKK}}
\end{minipage}
\end{figure}

Fig.~\ref{fig:JKK} shows the colour-magnitude diagram of our target Miras.
OGLE Miras in the LMC \citep{Soszynski-2009} are compared 
after a vertical offset corresponding to the approximate difference
in distance modulus,
$\Delta\mu _0=4$~mag, between the bulge and the LMC.
The line-of-sight depth of the bulge is much larger than
that of the LMC
and the colours and magnitudes of the bulge stars
are affected by interstellar extinction, while the LMC Miras
have small interstellar reddenings.
The very bright Mira is No.~06 which was classified 
as a C-rich star based on its IRAS/LRS spectrum \citep{Groenewegen-1993}.
This is obviously a foreground object.
\citet{Whitelock-2006} estimated its distance as 1.40~kpc and
foreground extinction, $A_V=0.80$~mag.
The other C-rich Miras indicated by filled circles are scattered around
a sequence roughly expected for such objects at around the distance
of the bulge.
Two of them look slightly brighter than the other Miras, and one
somewhat fainter; we discuss their distances further below.

The C-rich stars from \citet{Azzopardi-1988,Azzopardi-1991} are also shown 
in Fig.~\ref{fig:JKK}.
These are faint for evolved C-rich AGB stars
at the distance of the bulge 
as previously known
\citep[e.g.][]{Azzopardi-1988,Whitelock-1993,Ng-1997}. 
The faint C-rich stars in the Sculptor dwarf spheroidal \citep{Menzies-2011} 
have $\MK < -5.1$, which would give $\Ks < 9.5$ at the distance 
of the Galactic bulge. All of the bulge Azzopardi stars are fainter than this, 
so plausibly assuming that they are not all more distant than 
the bulge, they cannot have been enriched in carbon through 
dredge up in a normal star.
They are also bluer than C-rich Miras
in the LMC and the ones newly found in the bulge
(note that the $J-\Ks$ colours in Fig.~\ref{fig:JKK} are not dereddened).
\citet{Ng-1997,Ng-1998} discussed the possibility that
some of these C-rich stars may belong to the Sagittarius dwarf galaxy,
but their natures remain unclear.

The star symbol in Fig.~\ref{fig:JKK} indicates
the C-rich star found by \citet{Hynes-2014}.
They identified this as the optical counterpart of an X-ray source,
CXOGES J173620.2$-$293338.
The position in the colour-magnitude diagram
is close to that of the objects of \citet{Azzopardi-1991}, but
bluer and fainter than our C-rich Miras.
It shows small-amplitude irregular variations in its OGLE light curve
\citep{Udalski-2012}.
Its nature is unclear but the association of X-ray emission makes
it particularly interesting and suggests that it is a binary source.

\begin{figure}
\begin{minipage}{83mm}
\begin{center}
\includegraphics[clip,width=0.95\hsize]{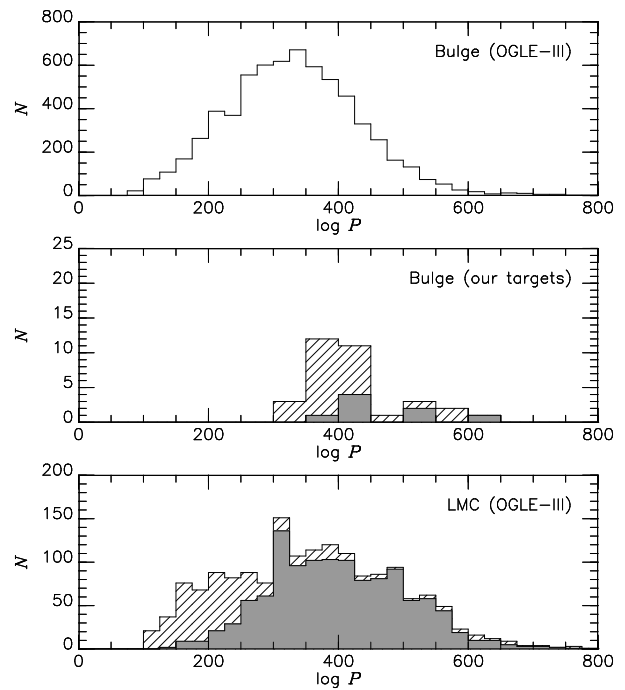}
\end{center}
\caption{
Period distributions of
OGLE-III bulge Miras \citep{Soszynski-2013} (top),
our targets (middle), and OGLE-III Miras in
the LMC \citep{Soszynski-2009} (bottom).
Except in the top panel, the filled and shaded portions indicate
C- and O-rich Miras respectively.
\label{fig:Phist}}
\end{minipage}
\end{figure}

In Fig.~\ref{fig:JKK},
the O-rich Miras we observed are indicated by open squares,
and most of them are
significantly bluer than the C-rich targets.
This suggests that the latter have intrinsically larger $J-\Ks$ as expected.
Compared to the O-rich Miras in the LMC (grey dots in Fig.~\ref{fig:JKK}),
our sample of O-rich Miras are redder, presumably due to the combined effect
of circumstellar and interstellar dust,
and at around the bright end of the LMC distribution.
Fig.~\ref{fig:Phist} shows that the spectroscopic targets we selected
are biased towards longer periods which explains
the higher luminosities.

\subsection{Distances}

Estimating accurate distances to the C-rich Miras is not straightforward.
It is known that reddish C-rich Miras appear fainter than
predicted by the near-infrared period-luminosity relation 
for bluer Miras. The deviation from the relation is correlated
with the colour in a quadratic manner \citep{Ita-2011};
such a correlation can be used
for correcting the luminosities and estimating the distances.
However, Miras towards the bulge are also reddened by 
interstellar dust. The interstellar extinction is represented by
the total-to-selective extinction ratio,
$A_{\Ks}/E_{J-\Ks}= 0.494$ \citep{Nishiyama-2006}, which is different from
the quadratic form of the deviation from the period-luminosity relation.
These two effects are mixed and difficult to separate.

\begin{figure}
\begin{minipage}{83mm}
\begin{center}
\includegraphics[clip,width=0.95\hsize]{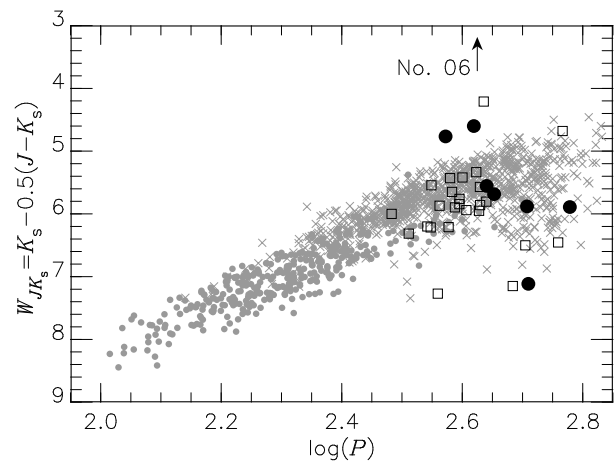}
\end{center}
\caption{
Period-Wesenheit relation.
Symbols are the same as in Fig.~\ref{fig:JKK}.
The bright foreground C-rich Mira, No.~06,
is outside the range of the plot but indicated by the arrow.
\label{fig:PWJK}}
\end{minipage}
\end{figure}

Here we consider the Wesenheit index,
$W_{J\Ks} = \Ks - 0.5 ( J-\Ks )$,
to make approximate corrections for the reddening and dimming.
The coefficient 0.5 is close to 
the total-to-selective extinction ratio and,
to some extent, the ratio between the colour and the deviation from
the period-luminosity relation, and thus $W_{J\Ks}$ is
insensitive to both reddening effects.
Fig.~\ref{fig:PWJK} plots the same objects, except non-Miras
as in Fig.~\ref{fig:JKK} on the $\log P$-$W_{J\Ks}$ diagram. 
Besides the clearly foreground object, No.~06, two stars (Nos.~01 and 03) are
distinctly brighter than the others. They are also bright in Fig.~\ref{fig:JKK}.
This suggests that the latter two are also foreground objects. 
Although it is difficult to estimate their distances as mentioned above,
they can be located at ${\sim}5$~kpc from the Sun
considering that they look brighter than the bulk of other Miras by
approximately 1~mag. This is around the short end of the distance
estimated by \citet{Miszalski-2013} for the symbiotic star No.~01.
Those authors also discussed the star's
possible membership of the extended
bar-like Galactic bulge, which we cannot rule out. The fact that star
No.~08 appears to be more distant than the bulk of the bulge stars may be
an indication that we are simply observing the large spread of the bulge
in the line of sight or it may indicate that this star is undergoing
an obscuration event of the type common in C-rich Miras \citep{Whitelock-2006}.
Future tests of bulge membership should be based on kinematic measurements,
both radial velocities from infrared high-resolution spectroscopy and
proper motion, e.g.~based on \textit{Gaia} and/or the VVV (VISTA Variables
in the Via Lactea) survey \citep{Libralato-2015}.

The colours and magnitudes of the other four C-rich Miras,
Nos.~02, 04, 05 and 07, 
are consistent with what are expected for C-rich Miras
at the distance of the bulge (Figs.~\ref{fig:JKK} and \ref{fig:PWJK}).

\section{Discussion}
\label{sec:discussion}

\subsection{Diagnostic to select C-rich Miras}

We have demonstrated that
the $(J-\Ks)$ vs $([9]-[18])$ diagram effectively isolates
candidate C-rich stars, although the separation between
C-rich and O-rich stars in that diagram is not definitive.
Colour variations and/or photometric errors
can increase the number of false positive candidates
as discussed in Section~\ref{sec:IRSF}.
In a stellar system like the Galactic bulge where O-rich objects are dominant,
it is hard to isolate individual C-rich objects.
Nevertheless, this method has proved useful enough to
find the new C-rich Miras in the bulge.
As discussed in Section~\ref{sec:phot}, C-rich Miras tend to show
cycle-to-cycle variations (Fig.~\ref{fig:LCs_C}).
The combination of light curve behaviour and the position in
the $(J-\Ks)$-$([9]-[18])$ diagram, or just the light curve behaviour alone,
might be a good diagnostic for selecting candidate C-rich Miras.

\subsection{The origin of C-rich Miras in the bulge}

The origin of the bulge C-rich Miras is difficult to establish.
We consider three possibilities below.

The five C-rich Miras (Nos.~03--05, 07, and 08)
are, within the uncertainties, in the bulge.
They have periods between 373 and 512~days,
and are not obviously interacting binaries.
If they have evolved from isolated stars they are of intermediate age,
0.5--3~Gyr, according to the correlation between age and period for
C-rich Miras \citep{Feast-2006}. This would imply the existence of
a small population of intermediate age stars with low metallicities,
or at least low oxygen abundances.
The presence of intermediate age stars in the bulge has long been suspected,
{e.g.}~from the presence there of longer period O-rich Miras 
and related objects (see \citealt{Catchpole-2016,Nataf-2016a} for reviews,
and among others on this general topic,
\citealt{vanLoon-2003,Groenewegen-2005}).
However, at a time when a case was being made for an entirely old bulge,
it was suggested
that these variables could be merged binaries \citep{Renzini-1990}. 
More recently evidence has been found for relatively metal-rich dwarfs
with ages of ${\sim 5}$~Gyr or younger
\citep{Bensby-2013,Haywood-2016}.
and for stars (of unknown age) with $\alpha$-element abundances
below the general bulge $\alpha$--Fe correlation \citep{RecioBlanco-2017}.
This serves to illustrate the complexity of the bulge population and
further exploration of this point is beyond the scope of this work.
If our C-rich Miras are intermediate age stars, this would be
the first demonstration that a (small) intermidiate age population exists
in the bulge with the correct mass, metallicity, and oxygen abundances to
produce C-rich stars in the AGB dredge-up process.

However, older stars can produce more massive, potentially C-rich, progeny through
either a mass transfer or a merger process. The C-rich Mira
in the globular cluster Lyng\r{a}~7 \citep{Feast-2013},
with a pulsation period of 551 days, is an example of a merger; the authors argued that the current mass of that Mira was about twice what was expected for a single Mira in that cluster. Such stars could be the descendants of blue stragglers, several of which have been clearly identified in the bulge \citep{Clarkson-2011}. It may also be that the Azzopardi C-rich stars are the giant branch and/or AGB predecessors of the C-rich Miras.

The two Miras likely in binary systems (Nos.~01 and 02, considering
their symbiotic nature) also have long periods, 416 and 600 days,
suggesting that they are binary mergers,
{i.e.} these were once triple systems,
or that they have accreted most of the mass of their companions. It is beyond
the scope of this paper to discuss the theoretical implications of this
in terms of stellar evolution or dynamics. It is certainly notable that
two out of the six C-rich Miras that are in or very near the bulge are binary.
Furthermore, the demonstration by \citet{Pietrzynski-2012}, that a bulge binary system involved an RR Lyr-like star with a mass of only  0.26~M$_\odot$,  shows that extreme mass transfer events, of the type that would be required to produce long period C-rich Miras, do occur.

A third alternative is that these C-rich Miras have been accreted from a merging dwarf galaxy with a composition similar to that of the Sagittarius dwarf galaxy, which contains numerous C-rich Miras in its core and in its tidal stream
\citep{Whitelock-1996,Law-2016}.

The merged binary scenario, or at least the scenario in which one star accretes most of the mass of a companion, is marginally preferred for the origin of these Miras, simply because mass-transfer binaries are found in the bulge and quite possibly become C-rich stars on the AGB. However,
in a very large and heterogeneous population such as the bulge it is impossible to be definitive about the origins of a small group of stars. Identification of more C-rich stars and studies of their kinematics and distribution will shed light on the problem. An abundances analysis would help a great deal, but that is not  practical for these very cool stars with extended
and dynamic atmospheres.

\section*{Acknowledgments}

We are grateful for
Lisa Crause's assistance with the new SpUpNIC spectrograph
on the first night of our observing run.
We thank Shazrene Mohamed for helpful discussions
about binary and multiple star systems.
This work has been supported
as a Joint Research Project under agreement between
the the Japan Society for the Promotion of Science (JSPS)
and National Research Foundation in South Africa.
NM is grateful to Grant-in-Aid (No.~26287028) from the JSPS. 
MWF, JWM, PAW, and EA acknowledge support from the NRF.
The IRSF project is a collaboration between Nagoya University and the
South African Astronomical Observatory (SAAO) supported by the
Grants-in-Aid for Scientific Research on Priority Areas (A)
(Nos.~10147207 and 10147214) and Optical \& Near-Infrared Astronomy
Inter-University Cooperation Program, from the Ministry of
Education, Culture, Sports, Science and Technology (MEXT) of Japan and
the National Research Foundation (NRF) of South Africa.

\bibliographystyle{mn2e}

\label{lastpage}

\end{document}

%% file: tab1.tex
\begin{table}
\begin{minipage}{80mm}
\caption{
List of targets we observed with the SpUpNIC spectrograph
and their types.
The ``Object'' column gives the OGLE ID numbers in the form of OGLE-BLG-LPV-NNNNNN except the three taken from \citet{Catchpole-2016} with the IDs of C16CCN.
\label{tab:spec}}
\begin{center}
\begin{tabular}{ccrrcrl}
\hline 
No. & Object & \multicolumn{1}{c}{$l$}  & \multicolumn{1}{c}{$b$} & UT & \multicolumn{1}{c}{Exp} & Type \\ 
 & & \multicolumn{1}{c}{$(\degr)$} & \multicolumn{1}{c}{$(\degr)$} & (in 2016) & \multicolumn{1}{c}{(s)} & \\ 
\hline 
01 & 149402 & $+2.0186$ & $-2.0559$ & 7/13 19:24 & 420 & C$^\dagger$ \\ 
02 & 145820 & $-2.7870$ & $-4.7249$ & 7/14 22:37 & 2400 & C \\ 
03 & 056745 & $-0.3282$ & $-1.2751$ & 7/13 22:09 & 600 & C \\ 
04 & 169921 & $-0.1137$ & $-3.7556$ & 7/13 21:03 & 600 & C \\ 
05 & 189627 & $+2.4710$ & $-2.9176$ & 7/14 20:23 & 1200 & C \\ 
06 & C16CC1 & $-9.9378$ & $-6.5831$ & 7/18 18:38 & 300 & C \\ 
07 & C16CC3 & $+7.3177$ & $-8.0958$ & 7/18 18:54 & 600 & C \\ 
08 & 230835 & $+7.1745$ & $-4.7850$ & 7/17 01:07 & 1200 & C: \\ 
09 & 194902 & $+2.2482$ & $-3.2756$ & 7/15 00:39 & 300 & M \\ 
10 & 013404 & $+0.1418$ & $2.5743$ & 7/13 20:17 & 420 & M \\ 
11 & 031232 & $+3.9599$ & $2.3630$ & 7/15 00:51 & 300 & M \\ 
12 & 039024 & $+4.7764$ & $2.4548$ & 7/18 18:23 & 300 & M \\ 
13 & 039561 & $+5.3812$ & $2.7934$ & 7/16 19:00 & 300 & M \\ 
14 & 110786 & $-2.3417$ & $-3.5774$ & 7/13 20:40 & 300 & M \\ 
15 & 122034 & $-1.8007$ & $-3.5504$ & 7/16 23:35 & 300 & M \\ 
16 & 127293 & $+0.2527$ & $-2.5135$ & 7/17 18:55 & 300 & M \\ 
17 & 131690 & $-0.3688$ & $-2.9776$ & 7/17 20:07 & 300 & M \\ 
18 & 133170 & $+0.1915$ & $-2.6888$ & 7/19 00:06 & 300 & M \\ 
19 & 170446 & $+1.8618$ & $-2.6492$ & 7/15 00:22 & 300 & M \\ 
20 & 188857 & $+0.0420$ & $-4.2511$ & 7/15 00:08 & 300 & M \\ 
21 & 195036 & $+1.8849$ & $-3.4856$ & 7/16 23:55 & 300 & M \\ 
22 & 195810 & $+2.0728$ & $-3.4196$ & 7/13 23:04 & 600 & M \\ 
23 & 214635 & $+5.1753$ & $-2.6767$ & 7/17 01:45 & 300 & M \\ 
24 & 216952 & $-0.1630$ & $-5.7607$ & 7/17 00:13 & 1200 & M \\ 
25 & 230250 & $+6.9510$ & $-4.2590$ & 7/16 21:18 & 120 & M \\ 
26 & 072230 & $-0.2513$ & $-1.5968$ & 7/14 18:19 & 900 & M \\ 
27 & 076743 & $+0.5348$ & $-1.2234$ & 7/13 21:38 & 600 & M \\ 
28 & 110674 & $-0.9474$ & $-2.7629$ & 7/19 00:27 & 300 & M \\ 
29 & 216273 & $+2.4655$ & $-4.2839$ & 7/16 20:12 & 600 & M \\ 
30 & 194903 & $-0.2322$ & $-4.6589$ & 7/16 22:01 & 600 & M \\ 
31 & 168224 & $+0.9524$ & $-3.1055$ & 7/17 20:23 & 300 & M \\ 
32 & 132552 & $+1.2480$ & $-2.0630$ & 7/16 19:33 & 1200 & M \\ 
33 & C16CC5 & $+5.7962$ & $-7.9836$ & 7/18 19:16 & 600 & M \\ 
34 & 082471 & $-0.3249$ & $-1.8431$ & 7/13 22:38 & 600 & --- \\ 
35 & 151973 & $+1.4302$ & $-2.4562$ & 7/15 01:29 & 360 & --- \\ 
36 & 179847 & $+0.8947$ & $-3.4740$ & 7/13 23:28 & 600 & --- \\ 
\hline 
\end{tabular}
\end{center}
$\dagger$\  No.~1 is the C-rich star (H1-45) identified by \citet{Miszalski-2013} as a symbiotic.
\end{minipage}
\end{table}

%% file: tab2.tex
\begin{table}
\begin{minipage}{80mm}
\caption{
Photometric data for target Miras.
When two magnitudes are given for each of the mean magnitudes ($\langle J\rangle$, $\langle H \rangle$, and $\langle \Ks \rangle$), the first one is obtained with the fitted light curves and the second one with simply averaging the $N$ measureents.
\label{tab:phot}}
\begin{center}
\begin{tabular}{lrcccc}
\hline 
No. & \multicolumn{1}{c}{Period} & \multicolumn{1}{c}{$N$} & \multicolumn{1}{c}{$\langle J\rangle$} & \multicolumn{1}{c}{$\langle H \rangle$} & \multicolumn{1}{c}{$\langle \Ks \rangle$} \\ 
 & \multicolumn{1}{c}{(days)} & & \multicolumn{1}{c}{(mag)} & \multicolumn{1}{c}{(mag)} & \multicolumn{1}{c}{(mag)} \\ 
\hline 
01 & 416.2 & 8 & 8.70/8.98 & 7.02/7.25 & 5.84/6.06 \\ 
02 & 600.5 & 7 & 12.35/12.37 & 9.95/9.96 & 8.04/8.05 \\ 
03 & 373.6 & 8 & 9.00/9.37 & 7.22/7.48 & 6.17/6.30 \\ 
04 & 449.7 & 6 & 10.73/10.53 & 8.93/8.74 & 7.47/7.30 \\ 
05 & 509.8 & 6 & 10.94/11.22 & 8.94/9.20 & 7.45/7.66 \\ 
06$^\dagger$ & 421.6 & --- & 8.41/--- & 5.79/--- & 3.95/--- \\ 
07$^\dagger$ & 436.9 & --- & 10.34/--- & 8.49/--- & 7.15/--- \\ 
08 & 512.6 & 7 & 13.07/13.16 & 11.29/11.04 & 9.37/9.13 \\ 
09 & 348.4 & 7 & 8.61/8.90 & 7.54/8.03 & 6.90/7.10 \\ 
10 & 365.1 & 7 & 8.17/8.60 & 6.98/7.57 & 6.21/6.78 \\ 
11 & 420.1 & 6 & 8.72/8.89 & 7.25/7.35 & 6.44/6.52 \\ 
12 & 436.5 & 9 & 8.88/9.30 & 7.51/7.86 & 6.67/6.97 \\ 
13 & 398.5 & 9 & 8.13/8.09 & 7.17/6.98 & 6.45/6.31 \\ 
14 & 432.1 & 8 & 7.90/7.87 & 6.38/6.40 & 5.38/5.43 \\ 
15 & 352.8 & 7 & 9.39/9.35 & 8.05/8.05 & 7.28/7.26 \\ 
16 & 393.8 & 7 & 8.70/8.89 & 7.48/7.67 & 6.69/6.86 \\ 
17 & 379.9 & 8 & 7.87/7.89 & 6.85/6.90 & 6.20/6.25 \\ 
18 & 424.6 & 6 & 8.60/8.77 & 7.48/7.67 & 6.75/6.89 \\ 
19 & 353.9 & 7 & 7.66/8.05 & 6.71/7.07 & 6.13/6.38 \\ 
20 & 304.0 & 6 & 8.55/8.49 & 7.50/7.48 & 6.83/6.83 \\ 
21 & 377.8 & 7 & 8.84/9.30 & 7.71/8.05 & 6.98/7.24 \\ 
22 & 426.0 & 6 & 8.58/8.93 & 7.36/7.62 & 6.51/6.69 \\ 
23 & 393.9 & 8 & 9.25/9.00 & 7.88/7.78 & 6.93/6.84 \\ 
24 & 574.4 & 7 & 10.30/10.76 & 8.81/9.18 & 7.55/7.89 \\ 
25 & 584.3 & 8 & 8.10/8.04 & 6.69/6.63 & 5.84/5.80 \\ 
26 & 506.6 & 8 & 10.22/11.00 & 8.65/9.42 & 7.34/8.00 \\ 
27 & 387.1 & 8 & 8.82/8.97 & 7.56/7.79 & 6.69/6.92 \\ 
28 & 426.2 & 8 & 9.30/9.40 & 7.92/8.01 & 6.98/7.04 \\ 
29 & 382.9 & 6 & 9.09/8.68 & 7.76/7.42 & 6.92/6.66 \\ 
30 & 362.6 & 7 & 10.97/10.63 & 9.70/9.39 & 8.61/8.39 \\ 
31 & 324.5 & 7 & 8.83/9.15 & 7.73/8.02 & 7.05/7.26 \\ 
32 & 482.8 & 7 & 10.69/11.17 & 9.21/9.59 & 8.21/8.49 \\ 
33$^\dagger$ & 404.4 & --- & 10.56/--- & 8.66/--- & 7.48/--- \\ 
34 & 434.9 & 7 & 10.18/10.47 & 8.40/8.64 & 7.04/7.21 \\ 
35 & 415.9 & 8 & 8.56/8.73 & 7.39/7.54 & 6.66/6.79 \\ 
36 & 486.8 & 18 & 12.28/12.46 & 10.13/10.00 & 8.20/8.18 \\ 
\hline 
\end{tabular}
\end{center}
$^\dagger$~$JHK$ magnitudes from \citet{Catchpole-2016}.
\end{minipage}
\end{table}

%% file: article.bbl
\begin{thebibliography}{99}
\bibitem[\protect\citeauthoryear{Alksnis {et~al.}}{2001}]{Alksnis-2001}
Alksnis A., Balklavs A., Dzervitis U., Eglitis I.,
Paupers O., Pundure I., 2001, Baltic Astron. 10, 1
\bibitem[\protect\citeauthoryear{Azzopardi {et~al.}}{1988}]{Azzopardi-1988}
{Azzopardi} M., {Lequeux} J., {Rebeirot} E., 1988, A\&A, 202, L27
\bibitem[\protect\citeauthoryear{Azzopardi {et~al.}}{1991}]{Azzopardi-1991}
Azzopardi M., Lequeux J., Rebeirot E., Westerlund B.~E., 1991, A\&AS, 88, 265
\bibitem[\protect\citeauthoryear{Beers \& Christlieb}{2005}]{Beers-2005}
Beers T.~C., Christlieb N., 2005, ARA\&A, 43, 531
\bibitem[\protect\citeauthoryear{Bensby {et~al.}}{2013}]{Bensby-2013}
Bensby T. et al., 2013, A\&A, 549, A147
\bibitem[\protect\citeauthoryear{Catchpole {et~al.}}{2016}]{Catchpole-2016}
Catchpole R.~M., Whitelock P.~A., Feast M.~W., Hughes S.~M.~G.,
Irwin M., Alard C., 2016, MNRAS, 455, 2216
\bibitem[\protect\citeauthoryear{Clarkson {et~al.}}{2011}]{Clarkson-2011}
Clarkson W.~I. et al., 2011, ApJ, 735, 37
\bibitem[\protect\citeauthoryear{Cole \& Weinberg}{2002}]{Cole-2002}
Cole A.~A., Weinberg M.~D., 2002, ApJ, 574, L43
\bibitem[\protect\citeauthoryear{Crause {et~al.}}{2016}]{Crause-2016}
Crause L. A. et al., 2016, Proc. SPIE, 9908, id. 990827
\bibitem[\protect\citeauthoryear{Feast \& Whitelock}{2000}]{Feast-2000}
Feast M.~W., Whitelock P. A., 2000, MNRAS, 317, 460
\bibitem[\protect\citeauthoryear{Feast}{2008}]{Feast-2008}
Feast M.~W., 2008, in First Middle East--Africa, Regional IAU Meeting. p3
Online at http://www.mearim.cu.edu.eg/new/Proceeding.htm
\bibitem[\protect\citeauthoryear{Feast, Whitelock, \& Menzies}{2006}]{Feast-2006}
Feast M.~W., Whitelock P.~A., Menzies J.~W., 2006, MNRAS, 369, 791
\bibitem[\protect\citeauthoryear{Feast {et~al.}}{2013}]{Feast-2013}
{Feast} M.~W., {Menzies} J.~W., {Whitelock} P.~A., 2013, MNRAS, 428, L36
\bibitem[\protect\citeauthoryear{Green}{2013}]{Green-2013}
Green P., 2013, ApJ, 765, 12
\bibitem[\protect\citeauthoryear{Green \& Margon}{1994}]{Green-1994}
Green P. J., Margon B., 1994, ApJ, 423, 723
\bibitem[\protect\citeauthoryear{Groenewegen \& Blommaert}{2005}]{Groenewegen-2005}
Groenewegen M.~A.~T., Blommaert J.~A.~D.~L., 2005, A\&A, 443, 143
\bibitem[\protect\citeauthoryear{Groenewegen, de Jong, \& Baas}{1993}]{Groenewegen-1993}
Groenewegen M.~A.~T., de Jong T., Baas F., 1993, A\&AS, 101, 513
\bibitem[\protect\citeauthoryear{Haywood {et~al.}}{2016}]{Haywood-2016}
Haywood M., Di Matteo P., Snaith O., Calamida A., 2016, A\&A, 593, A82
\bibitem[\protect\citeauthoryear{Huxor \& Grebel}{2015}]{Huxor-2015}
Huxor, A.~P., Grebel, E.~K., 2015, MNRAS, 453, 2653
\bibitem[\protect\citeauthoryear{Hynes {et~al.}}{2014}]{Hynes-2014}
Hynes R.~I. et al., 2014, ApJ, 780, 11
\bibitem[\protect\citeauthoryear{Ishihara {et~al.}}{2010}]{Ishihara-2010}
Ishihara D. et al., 2010, A\&A, 514, A1
\bibitem[\protect\citeauthoryear{Ishihara {et~al.}}{2011}]{Ishihara-2011}
Ishihara D., Kaneda H., Onaka T., Ita Y., Matsuura M., Matsunaga N., 2011, A\&A, 534, A79
\bibitem[\protect\citeauthoryear{Ita {et~al.}}{2010}]{Ita-2010}
Ita Y. et al., 2010, A\&A, 514, A2
\bibitem[\protect\citeauthoryear{Ita \& Matsunaga}{2011}]{Ita-2011}
Ita Y., Matsunaga N., 2011, MNRAS, 412, 2345
\bibitem[\protect\citeauthoryear{Izzard, Jeffery, \& Lattanzio}{2007}]{Izzard-2007}
{Izzard} R.~G., {Jeffery} C.~S., {Lattanzio} J., 2007, A\&A, 470, 661
\bibitem[\protect\citeauthoryear{Lan\c{c}on \& Wood}{2000}]{Lancon-2000}
Lan\c{c}on A., Wood P.~R., 2000, A\&AS, 146, 217
\bibitem[\protect\citeauthoryear{Law \& Majewski}{2016}]{Law-2016}
Law D.~R., Majewski S.~R., 2016, ASSL, 420, 31
\bibitem[\protect\citeauthoryear{Libralato {et al.}}{2015}]{Libralato-2015}
Libralato M. et al., 2015, MNRAS, 450, 1664
\bibitem[\protect\citeauthoryear{Loidl, Lan\c{c}on, \& J{\o}rgensen}{2001}]{Loidl-2001}
Loidl R., Lan\c{c}on A., J{\o}rgensen U.~G., 2001, A\&A, 371, 1065
\bibitem[\protect\citeauthoryear{Marigo {et~al.}}{2008}]{Marigo-2008}
Marigo P., Girardi L., Bressan A., Groenewegen M.~A.~T., Silva L., Granato G.~L., 2008, A\&A, 482, 883
\bibitem[\protect\citeauthoryear{McClure \& Woodsworth}{1990}]{McClure-1990}
McClure R.~D., Woodsworth A.~W., 1990, ApJ, 352, 709
\bibitem[\protect\citeauthoryear{McWilliam}{2016}]{McWilliam-2016}
McWilliam A. 2016, PASA, 33, e040
\bibitem[\protect\citeauthoryear{Menzies {et~al.}}{2011}]{Menzies-2011}
Menzies J.~W., Feast M.~W., Whitelock P.~A., Matsunaga N., 2011, MNRAS, 414, 3492
\bibitem[\protect\citeauthoryear{Miszalski {et~al.}}{2013}]{Miszalski-2013}
Miszalski B., Miko{\l}ajewska J., Udalski A., 2013, MNRAS, 432, 3186
\bibitem[\protect\citeauthoryear{Mouhcine \& Lan\c{c}on}{2003}]{Mouhcine-2003}
Mouhcine M., Lan\c{c}on A., 2003, MNRAS, 338, 572
\bibitem[\protect\citeauthoryear{Nagashima {et~al.}}{1999}]{Nagashima-1999}
Nagashima, C. {et~al.}, 1999, in Nakamoto T., ed., Proc. Star Formation 1999. Nobeyama Radio Observatory, Nagano, p.~397
\bibitem[\protect\citeauthoryear{Nagayama {et~al.}}{2003}]{Nagayama-2003}
Nagayama T. {et~al.}, 2003, in Iye M., Moorwood A.~F.~M., eds, Proc. SPIE Vol.~4841, Instrument Design and Performance for Optical/Infrared Ground-based Telescopes. SPIE, Bellingham, p.~459
\bibitem[\protect\citeauthoryear{Nataf}{2016a}]{Nataf-2016a}
Nataf D.~M., 2016a, PASA, 33, e023
\bibitem[\protect\citeauthoryear{Nataf {et~al.}}{2016b}]{Nataf-2016b}
Nataf D.~M. {et~al.}, 2016, MNRAS, 456, 2692
\bibitem[\protect\citeauthoryear{Ness \& Freeman}{2016}]{Ness-2016}
Ness M., Freeman K., 2016, PASA, 33, 22
\bibitem[\protect\citeauthoryear{Ng}{1997}]{Ng-1997}
Ng Y.~K., 1997, A\&A, 328, 211
\bibitem[\protect\citeauthoryear{Ng}{1998}]{Ng-1998}
Ng Y.~K., 1998, A\&A, 338, 435
\bibitem[\protect\citeauthoryear{Nishiyama {et~al.}}{2006}]{Nishiyama-2006}
Nishiyama S. et al., 2006, ApJ, 638, 839
\bibitem[\protect\citeauthoryear{Ojha {et~al.}}{2007}]{Ojha-2007}
Ojha D.~K., Tej A., Schultheis M., Omont A., Schuller F., 2007, MNRAS, 381, 1219
\bibitem[\protect\citeauthoryear{Pietrzy\'nski {et~al.}}{2012}]{Pietrzynski-2012}
Pietrzy\'nski G., et al., 2012, Nature, 484, 75
\bibitem[\protect\citeauthoryear{Plant {et~al.}}{2016}]{Plant-2016}
Plant K.~A., Margon B., Guhathakurta P., Cunningham E.~C., Toloba E., Munn J.~A., 2016, ApJ, 833, 232
\bibitem[\protect\citeauthoryear{Recio-Blanco {et~al.}}{2017}]{RecioBlanco-2017}
Recio-Blanco A. et al., 2017, A\&A, in press; arXiv:1702.04500 
\bibitem[\protect\citeauthoryear{Renzini \& Greggio}{1990}]{Renzini-1990}
Renzini, A., Greggio, L., 1990, in Jarvis B.~J., Terndrup D.~M., eds, Bulges of Galaxies. ESO, Garching, p.~47
\bibitem[\protect\citeauthoryear{Skrutskie {et~al.}}{2006}]{Skrutskie-2006}
Skrutskie M.~F. et al., 2006, AJ, 131, 1163
\bibitem[\protect\citeauthoryear{Soszy\'nski {et~al.}}{2009}]{Soszynski-2009}
Soszy\'nski I. et al., 2009, AcA, 59, 239
\bibitem[\protect\citeauthoryear{Soszy\'nski {et~al.}}{2013}]{Soszynski-2013}
Soszy\'nski I. et al., 2013, AcA, 63, 21
\bibitem[\protect\citeauthoryear{Totten, Irwin, \& Whitelock}{2000}]{Totten-2000}
Totten E.~J., Irwin M.~J., Whitelock P.~A., 2000, MNRAS, 314, 630
\bibitem[\protect\citeauthoryear{Tyson \& Rich}{1991}]{Tyson-1991}
Tyson N.~D., Rich R.~M., 1991, ApJ, 367, 547
\bibitem[\protect\citeauthoryear{Udalski {et~al.}}{2012}]{Udalski-2012}
Udalski A. et al., 2002, AcA, 62, 133
\bibitem[\protect\citeauthoryear{Uttenthaler {et~al.}}{2015}]{Uttenthaler-2015}
Uttenthaler S., Blommaert J.~A.~D.~L., Wood P.~R., Lebzelter T.,
Aringer B., Schultheis M., Ryde N., 2015, MNRAS, 451, 1750
\bibitem[\protect\citeauthoryear{van Loon {et~al.}}{2003}]{vanLoon-2003}
van Loon, J.~Th. et al. 2003, MNRAS, 338, 857
\bibitem[\protect\citeauthoryear{Vassiliadis \& Wood}{1993}]{Vassiliadis-1993}
Vassiliadis E., Wood P.~R., 1993, ApJ, 413, 641
\bibitem[\protect\citeauthoryear{Whitelcok}{1993}]{Whitelock-1993}
Whitelock P.~A., 1993, IAUS, 153, 39
\bibitem[\protect\citeauthoryear{Whitelock, Irwin, \& Catchpole}{1996}]{Whitelock-1996}
{Whitelock} P.~A., {Irwin} M., {Catchpole} R.~M., 1996, New Astron., 1, 57
\bibitem[\protect\citeauthoryear{Whitelock {et~al.}}{2003}]{Whitelock-2003}
Whitelock P.~A., Feast M.~W., van Loon J.~Th., Zijlstra A.,
2003, MNRAS, 342, 86
\bibitem[\protect\citeauthoryear{Whitelock {et~al.}}{2006}]{Whitelock-2006}
Whitelock P.~A., Feast M.~W., Marang F., Groenewegen M.~A.~T., 2006, MNRAS, 369, 751
\bibitem[\protect\citeauthoryear{Wright {et~al.}}{2009}]{Wright-2009}
Wright N.~J., Barlow M.~J., Greimel R., Drew J.~E., Matsuura M., Unruh Y.~C., Zijlstra A.~A., 2009, MNRAS, 400, 1413
\bibitem[\protect\citeauthoryear{Wood, Habing, \& McGregor}{1998}]{Wood-1998}
{Wood} P.~R., {Habing} H.~J., {McGregor} P.~J., 1998, A\&A, 336, 925
\bibitem[\protect\citeauthoryear{Wyckoff}{1970}]{Wyckoff-1970}
Wyckoff S., 1970, ApJ, 162, 203
\end{thebibliography}
